\documentclass [11pt,a4paper]{article}
\pdfoutput=1 
\usepackage{jheppub}
\usepackage{latexsym,epsfig,amssymb, amsmath,nicefrac}

\usepackage{color}
\usepackage[makeroom]{cancel}
  \usepackage{latexsym}
  \usepackage{epsf}
  \usepackage{amssymb}
  \usepackage{graphicx}
  \usepackage{amsmath}
  \usepackage{amsmath,amssymb,amsthm}
  \usepackage{verbatim}
  \usepackage{hyperref}

\newcommand{\e}{\textrm{e}}

\def\tb0{\tilde{\beta}_0}
{\def\b0{\beta_0}

\def\bi{\begin{itemize}}
\def\ei{\end{itemize}}
\def\be{\begin{equation}}
\def\ee{\end{equation}}
\newcommand{\bea}{\begin{eqnarray}}
\newcommand{\eea}{\end{eqnarray}}
 
\renewcommand{\Im}{\textrm{Im}\,}
\renewcommand{\Re}{\textrm{Re}\,}

\def\Kahler{K\"{a}hler~}

\begin{document}

\vspace{1cm}

\title{Cosmological  $\alpha$-Attractors and de Sitter Landscape}

\author{Marco Scalisi}
\emailAdd{m.scalisi@rug.nl}
\affiliation{Van Swinderen Institute for Particle Physics and Gravity, University of Groningen, \\ Nijenborgh 4, 9747 AG Groningen, The Netherlands}

\abstract{

We provide a unified description of cosmological $\alpha$-attractors and late-time acceleration, in excellent agreement with the latest Planck data. Our construction involves two superfields playing distinctive roles: one is the dynamical field and its evolution determines inflation and dark energy, the other is nilpotent and responsible for a landscape of vacua and supersymmetry breaking. We prove  that the attractor nature of the theory is enhanced when combining the two sectors:  cosmological attractors are  very stable with respect to any possible value of the cosmological constant and, interestingly, to any generic coupling of the inflationary sector with the field responsible for uplifting. Finally, as related result,  we show how specific couplings generate an arbitrary inflaton potential in a supergravity framework with varying \Kahler curvature.}

\maketitle

\section{Introduction}

Observational evidence \cite{Riess:1998cb,Perlmutter:1998np,Hinshaw:2012aka,Planck:2015xua,Ade:2015lrj}  seems to point at acceleration as a fundamental ingredient of our Universe. Primordial inflation is the leading paradigm to account for the origin of the anisotropies in the CMB radiation and, then, the formation of large scale structures. These are currently observed to experience a mysterious accelerating phase, whose source has been generically called dark energy. Although the origin of both early- and late-time acceleration still represents a great theoretical puzzle, the simple assumption that the potential energy of a scalar field may serve as fundamental source has turned out to be successful in terms of investigation, extraction of predictions and agreement with the present observational data. In the simplest scenario, a scalar field slowly rolls down along its potential, driving inflation, and eventually sits in a minimum with a small positive cosmological constant of the order $\Lambda\sim10^{-120}$ ($M_{Pl}=1$).

The embedding into high-energy physics frameworks, such as supergravity or string theory, seems to be natural. On the one hand, the high energy-scale of inflation would require UV-physics control. On the other hand, the anthropic argument in a landscape of many string vacua \cite{Linde:1986fd,Weinberg:1987dv, Bousso:2000xa,Kachru:2003aw,Douglas:2003um,Susskind:2003kw} would provide a possible explanation of the smallness of the current cosmological constant.

The most economical models of inflation in supergravity are those employing just one chiral superfield $\Phi$ \cite{Goncharov:1983mw,Goncharov:1985yu,Ketov:2014qha,Ketov:2014hya,Linde:2014ela,Linde:2014hfa,Roest:2015qya,Linde:2015uga}. In the simplest case of a canonical \Kahler potential and a general superpotential, one has
\be\label{K&Tsingle}
K=-\tfrac{1}{2}\left(\Phi-\bar{\Phi}\right)^2\,,\qquad W=f(\Phi)\,,
\ee
where $f$ is a real holomorphic function of its argument. The shift-symmetry in $K$ allows to identify the real part of $\Phi$ as the inflaton, since it is naturally light \cite{Kawasaki:2000yn}. In this case, the scalar potential  reads\footnote{The potential \eqref{Vsingle} is generated by means of the superpotential which sofly breaks the shift-symmetry in a natural way \cite{'tHooft:1979bh}. Furthermore, whereas higher-order corrections might yield small deformations in the inflationary dynamics (see e.g. \cite{Lyth:1998xn,Allahverdi:2006we}), we focus on the lower-order terms in order to discuss the main results of the paper.}
\be\label{Vsingle}
V=f'(\Phi)^2 -3f(\Phi)^2\,,
\ee
along $\Phi=\bar{\Phi}$, which is a consistent truncation. Then, one immediately faces the first problem when trying to implement inflation within a single superfield model: the two opposite sign contributions in Eq.~\eqref{Vsingle}  highly restrict the possibilities of yielding a positive scalar potential along the whole trajectory. In particular, the negative term can give rise to dangerous contributions spoiling the inflationary behavior at large field range. Several solutions have been proposed in order to overcome this issue, one of the most famous being introducing a second chiral field which appears linearly in $W$ and kills the negative term in the scalar potential \cite{Kawasaki:2000yn,Kallosh:2010xz}. Another recent solution consists in introducing higher order terms in $K$ in order to enhance the positive definite contribution in $V$, with no need to add  a second superfield \cite{Ketov:2014qha,Ketov:2014hya}. However, this scenario does not lead to a pure single-field truncation as a two fields dynamics generically appears near the minimum.

Curiously, the very first model of chaotic inflation in supergravity \cite{Goncharov:1983mw,Goncharov:1985yu,Linde:2014hfa} made already use  of the minimal setup with a single superfield having a canonical \Kahler potential (an embedding in curved \Kahler geometry is given in \cite{Kallosh:2015lwa}) but it did not suffer from the issue described above. In fact, the specific choice of the superpotential leads to a perfect balance between the positive and negative contributions in the scalar potential.  Recently, it has been pointed out that this scenario is just an example of a much larger class of inflationary models  \cite{Roest:2015qya}, with observational predictions given by
\begin{align}\label{nsandr}
n_s= 1-\frac{2}{N}\,,\qquad r=\frac{12\alpha}{N^2}\,,
\end{align}
with $N$ being the number of e-folds between horizon exit and the end of inflation. In \cite{Roest:2015qya}, it was shown that this class arises as a flat singular limit of a supergravity scenario where the geometric properties of the curved \Kahler manifold determines the observational predictions \eqref{nsandr}, for generic choices of the superpotential. In particular, the amount of primordial gravitational waves is directly related to the \Kahler curvature which depends on the parameter $\alpha$. This geometric phenomenon had been earlier discovered in \cite{Kallosh:2013hoa,Ferrara:2013rsa,Kallosh:2013yoa,Kallosh:2014rga,Galante:2014ifa} and dubbed as {\it cosmological $\alpha$-attractors}. Curiously, whereas the inflationary predictions of the geometric attractors are fully encoded in $K$, the spectral tilt and the tensor-to-scalar ratio of the models with a flat \Kahler geometry are determined by $W$. Interestingly, the attractor-structure is maintained also in the flat case  as, starting from a $W$ corresponding to a pure de Sitter phase,  just the first generic correction will determine \eqref{nsandr}, while higher order terms will be irrelevant. The mechanism is described in detail in \cite{Roest:2015qya} and throughout the present paper.

Uplifting the supersymmetric Minkowski minimum of these scenarios would be the natural next step in order to consider the current acceleration. However, it has been pointed out that obtaining a de Sitter vacuum from a SUSY one is subject to a number of restrictions \cite{Kallosh:2014oja}  which make a unified picture of inflation and dark energy very challenging to achieve, when using just one chiral superfield \cite{Linde:2014ela}. A way to overcome this issue is to add a nilpotent superfield $S$ \cite{Volkov:1972jx,Volkov:1973ix,Rocek:1978nb,Ivanov:1978mx,Lindstrom:1979kq,Casalbuoni:1988xh,Komargodski:2009rz}, defined as
\be\label{nilpo}
S^2(x,\theta)=0\,,
\ee
with $x$ and $\theta$ being respectively the bosonic and fermionic coordinates. In fact, the nilpotent field seems to be naturally related to de Sitter vacua and it has been used in order to construct  inflationary models with de Sitter exit and controllable level of SUSY breaking at the minumum \cite{Antoniadis:2014oya,Ferrara:2014kva,KLnil,Kallosh:2014wsa,Dall'Agata:2014oka,Kallosh:2014hxa,Kallosh:2015lwa} (a pure de Sitter supegravity theory has been recently constructed by means of a local version of the nilpotent multiplet \cite{Bergshoeff:2015tra}) . The two sectors appearing in these constructions have independent roles: the $\Phi$-sector contains the scalar which evolves and dynamically determines inflation and dark energy while the field $S$ is responsible for the landscape of vacua. However, in general, the inflationary regime is really sensitive to the coupling between the two sectors and to the value of the uplifting; as we will show in Sec.~\ref{InflDS}, one needs to make specific choices for the superpotential.

In this paper, we present special stability of $\alpha$-attractors when combined with a nilpotent sector. We prove that their inflationary predictions are extremely stable with respect to any possible value of the cosmological constant and to any generic coupling between $\Phi$ and $S$, exhibiting attractor structure also in the uplifting sector. These scenarios simply emerges as the most generic expansion of the superpotential. We will firstly show these results in Sec.~\ref{FlatUp} in the context of {\it flat $\alpha$-attractors}. Then, thanks to the correspondence introduced in \cite{Roest:2015qya}, we will show that analogous results hold also in the case of {\it geometric $\alpha$-attractors} with curved \Kahler geometry. Eventually, in the last section, we prove that  an arbitrary inflationary potential with controllable level of dark energy and SUSY breaking can be obtained even in a supergravity context with curved \Kahler manifold, independently of the value of the curvature. However, in this case, the appealing attractor structure of the model is lost as the coupling between $\Phi$ and $S$ must be specific. 

\section{Inflation and de Sitter exit} \label{InflDS}

In \cite{Kallosh:2014hxa} a general class of inflationary models with de Sitter exit and controllable level of SUSY breaking at the minimum was proposed. This is constructed by means of two chiral superfields $\Phi$ and $S$, the latter being nilpotent. The \Kahler potential and superpotential are of the form

\be\label{K&W}
K=-\tfrac{1}{2}\left(\Phi-\bar{\Phi}\right)^2+S\bar{S}\,,\quad W=f(\Phi)+ g(\Phi) S\,,
\ee
where $f$ and $g$ are real holomorphic functions of their arguments and $W$ has the the most general form, provided $S$ is nilpotent. The nilpotency condition \eqref{nilpo} on the field $S$  translates into replacing the scalar part of the supermultiplet with a bilinear combination of fermions. Then, in order to study the dynamical evolution of the system, one needs simply to declare that its vev vanishes without need to care for its stability.

Within this class of models, the real part of the field $\Phi$ plays the role of the inflaton, rolling down along $S=0$ and $\Phi=\bar{\Phi}$, and drives a potential which reads
\be\label{pot}
V=g(\Phi)^2+f'(\Phi)^2 -3f(\Phi)^2\,.
\ee
Note that the last two terms are exactly the ones appearing in \eqref{Vsingle}, that is, for a single superfield model \cite{Achucarro:2012hg,Roest:2013aoa}.

After inflation, the journey of $\Re\Phi$ ends into a minimum placed at $\Phi=0$, provided the functions $f$ and $g$ satisfy
\begin{align}\label{derivf&g}
f'(0)=g'(0)=0\,.
\end{align}

The values of $f$ and $g$ at the minimum will allow for a wide spectrum of possibilities in terms of supersymmetry breaking and cosmological constant, along the lines of the string landscape scenario. Supersymmetry is spontaneously broken just in the nilpotent direction\footnote{This allows for a simplification of the fermionic sector of the supergravity action. Specifically, in the unitary gauge, the gravitino interacts just with the fermion of the nilpotent field leading to a simple version of the super-Higgs mechanism \cite{Dall'Agata:2014oka,Kallosh:2014hxa}.}, namely
\begin{align}
D_S W_{min}=g(0)=M\,, \qquad D_\Phi W_{min}=0\,,
\end{align}
where we have introduced $M$ as SUSY breaking parameter. Further, the gravitino mass is given by $m_{3/2}=f(0)$. The value of the cosmological constant is equal to
\be\label{Vminimum}
\Lambda= g^2(0) - 3 f^2(0)=M^2-3m_{3/2}^2\,.
\ee

However, the generality of Eq.~\eqref{pot} does not assure always a viable inflationary scenario, for the same reasons described in the Introduction. The negative term can be dominating at large value of the inflaton field and not give rise to inflation. In the framework defined by Eq.~\eqref{K&W}, a successful choice for the functions  $f$ and $g$ is given by \cite{Dall'Agata:2014oka,Kallosh:2014hxa}
\be\label{g&f}
f(\Phi)=\beta\ g(\Phi)\,,
\ee
with $\beta$ being some constant. The specific relation \eqref{g&f} leads to a situation where the negative contribution in \eqref{pot} is exactly canceled when the minimum \eqref{Vminimum} is Minkowski and, then, by fine-tuning $\beta=1/\sqrt{3}$. Then, the scalar potential turns out to have the simple form $V=\left[f'(\Phi)\right]^2$. Allowing for a small cosmological constant $\Lambda\sim10^{-120}$ (then, having a tiny deviation of $\beta$ from $1/\sqrt{3}$) does not change effectively the inflationary predictions. Other possible choices for $f$ and $g$ are discussed in \cite{KLnil,Kallosh:2014hxa}.

This construction is quite flexible in terms of  observational predictions allowing for any possible value of $n_s$ and $r$. Nonetheless, the generality of such construction relies on the relation \eqref{g&f} and turns out to be really sensitive with respect to any other generic coupling between the inflaton and the nilpotent sector.  Moreover, the negative contribution of Eq.~\eqref{pot} is balanced just if one assumes the observational evidence of a negligible cosmological constant. A generic de Sitter landscape would yield important corrections to such construction.

\section{Uplifting flat $\alpha$-attractors} \label{FlatUp}

In the single superfield framework defined by \eqref{K&Tsingle}, inflationary models with observational predictions given by \eqref{nsandr} and in excellent agreement with Planck were found in \cite{Roest:2015qya}. These are defined by 
\be\label{fflat}
f(\Phi)= \e^{\sqrt{3} \Phi} - \e^{-\sqrt{3} \Phi} F \left(\e^{- 2 \Phi / \sqrt{3 \alpha}}\right) \,,
\ee
where $F$ is an arbitrary function having an expansion such as $F(x)=\sum_n c_n x^n$ with
\be\label{xflat}
x\equiv\e^{- 2 \Phi / \sqrt{3 \alpha}}\,.
\ee

This class of models, being characterized by exponentials as building blocks of the superpotential, manifestly exhibits its attractor nature through the insensitivity to the structure of $F$. While the constant term $c_0$ would yield a de Sitter plateau $V=12c_0$, the first linear term would define the inflationary fall-off typical of $\alpha$-attractors, such as
\be\label{falloff}
V=V_0+V_1 \e^{- \sqrt{\frac{2}{3\alpha}} \varphi }+...\,,
\ee
at large values of the canonical  scalar field $\varphi= \sqrt{2}\ \Re\Phi$, with $V_0=12c_0$ and $V_1=16c_1$, the latter being negative. Higher order terms would be unimportant for observational predictions. 

This scenario can be naturally embedded in the construction discussed in the previous section. A first step would be simply choosing \eqref{fflat} as function $f$ in Eq.~\eqref{K&W}. In fact, this represents a valid alternative to the specific choice \eqref{g&f}: it yields always a balance of the negative term in \eqref{pot}, independently of the value of the uplifting at the minimum, and, interestingly, it decouples the functional forms of $f$ and $g$.  As second step, one may notice that, given the form of the scalar potential Eq.~\eqref{pot}, any generic expansion such as
\be\label{f&gflat}
f(x)=\sum_n a_n x^n\,,\qquad g(x)=\sum_n b_n x^n\,,
\ee
with $x$ given by Eq.~\eqref{xflat}, would give rise to a fall-off from de Sitter analogous to Eq.~\eqref{falloff} with
\be \label{V0V1}
V_0 =b_0^2-3a_0^2\,,\qquad V_1=2b_0b_1-6a_0a_1\,,
\ee
and, then, yield the universal predictions \eqref{nsandr}.

It is remarkable that the attractor structure of the theory is enhanced when combining the inflaton  with the nilpotent sector. The inflationary regime is very stable with respect to any deformation of the superpotential and any value of the uplifting.

Within this construction, the condition \eqref{derivf&g} of a minimum placed at $\Phi=0$ ($x=1$) translates into 
\be\label{condmin}
\sum_{n=1} n\,a_n=0\,, \qquad \sum_{n=1} n\ b_n=0\,.
\ee

Interestingly, the value of the cosmological constant at the minimum is given by 
\be\label{CC}
\Lambda= \left(\sum_{n} b_n\right)^2-3\left(\sum_{n} a_n\right)^2\,,
\ee
and then as a sum of the coefficients of the expansions \eqref{f&gflat} which, separately, determine the gravitino mass and the scale of supersymmetry breaking, such as
\be\label{M32M}
m_{3/2}=\sum_{n} a_n\,, \qquad M=\sum_{n} b_n\,.
\ee

Stability of the inflationary regime in the imaginary direction is always assured, for any value of $\alpha$, as the condition is simply
\be\label{massIm}
|b_0|>|a_0|\,.
\ee
In fact, the mass of $\Im\Phi$ turns out to have a natural expansion at small value of $x$ (large values of $\varphi$) such as
\be\label{MassImaginary}
m_{\Im\Phi}^2= 2 (b_0^2-a_0^2) + \frac{4}{3\alpha} \left[b_0b_1(3\alpha -1) - a_0 a_1 (3\alpha +1) \right] x + ... \,,
\ee
that is, a deviation from a constant plateau typical of the scalar potential of $\alpha$-attractors, where  higher order terms do not play any role. During inflation, the $\Re \Phi$ moves along a valley of constant width. This phenomenon can be appreciated below in Fig.~\ref{Masses}, for a specific example.  Stability at the minimum is model dependent since, generically, the infinite tower of coefficients $a_n$ and $b_n$ contribute to the masses.

\begin{figure}[htb]
\hspace{-3mm}
\begin{center}
\includegraphics[width=7.5cm]{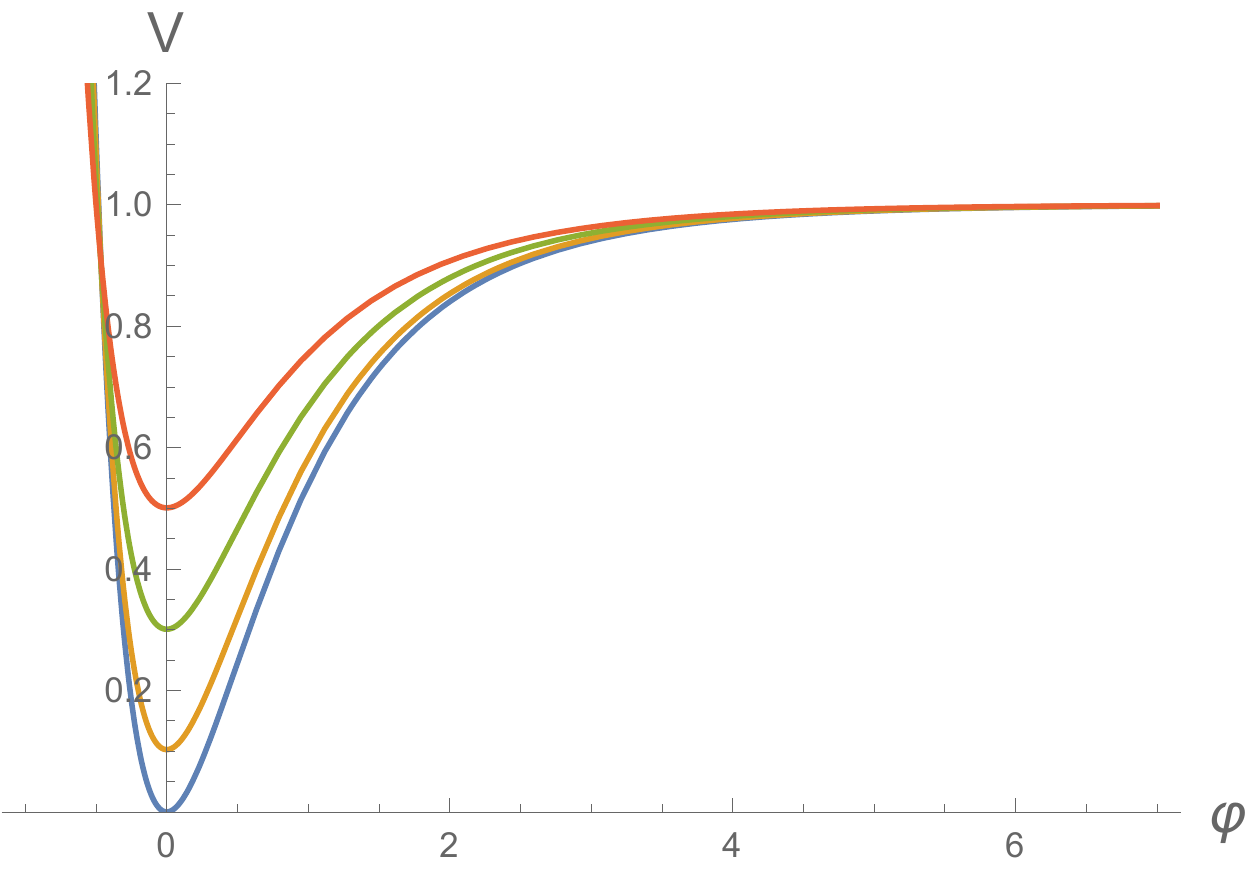}
\caption{Scalar potential of the model defined by Eq.~\eqref{Example} with $\alpha=1$ and uplifting equal to $\Lambda=\{0, 0.1, 0.3, 0.5\}.$}\label{Potential}
\end{center}
\vspace{0cm}
\end{figure}

The simplest example of such class of models is given by the following choice:
\be\label{Example}
f=a_0+a_1 x+a_2 x^2\,,\qquad g=b_0\,.
\ee

In fact, this is a minimum in order to have a deviation from de Sitter typical of $\alpha$-attractors, which comes from the linear term, and a non-trivial solution of Eq.~\eqref{condmin} to have a minimum placed at the origin, thanks to the quadratic contribution. Higher order terms will not affect neither the inflationary energy  nor the characteristic fall-off, as it is clear from Eq.~\eqref{V0V1}. The scalar potential, for $\alpha=1$ and different amount of uplifting, is shown in Fig.~\ref{Potential}. Stability occurs along the full inflationary trajectory and also at the minimum where both directions of $\Phi$ turn out to be stable, as it is shown in Fig.~\ref{Masses}. Analogous results hold for other values of $\alpha$.

The addition of higher order terms both in $f$ and $g$ would allow for more flexibility in terms of separation of the physical scales. In fact, whereas the inflationary regime would be absolutely insensitive to high order contributions, the coefficients of these terms turn out to be fundamental in determining the scale of SUSY breaking, the gravitino mass and the cosmological constant, as given by Eq.~\eqref{CC} and Eq.~\eqref{M32M}.

\begin{figure}[htb]
\hspace{-3mm}
\begin{center}
\includegraphics[width=7.5cm]{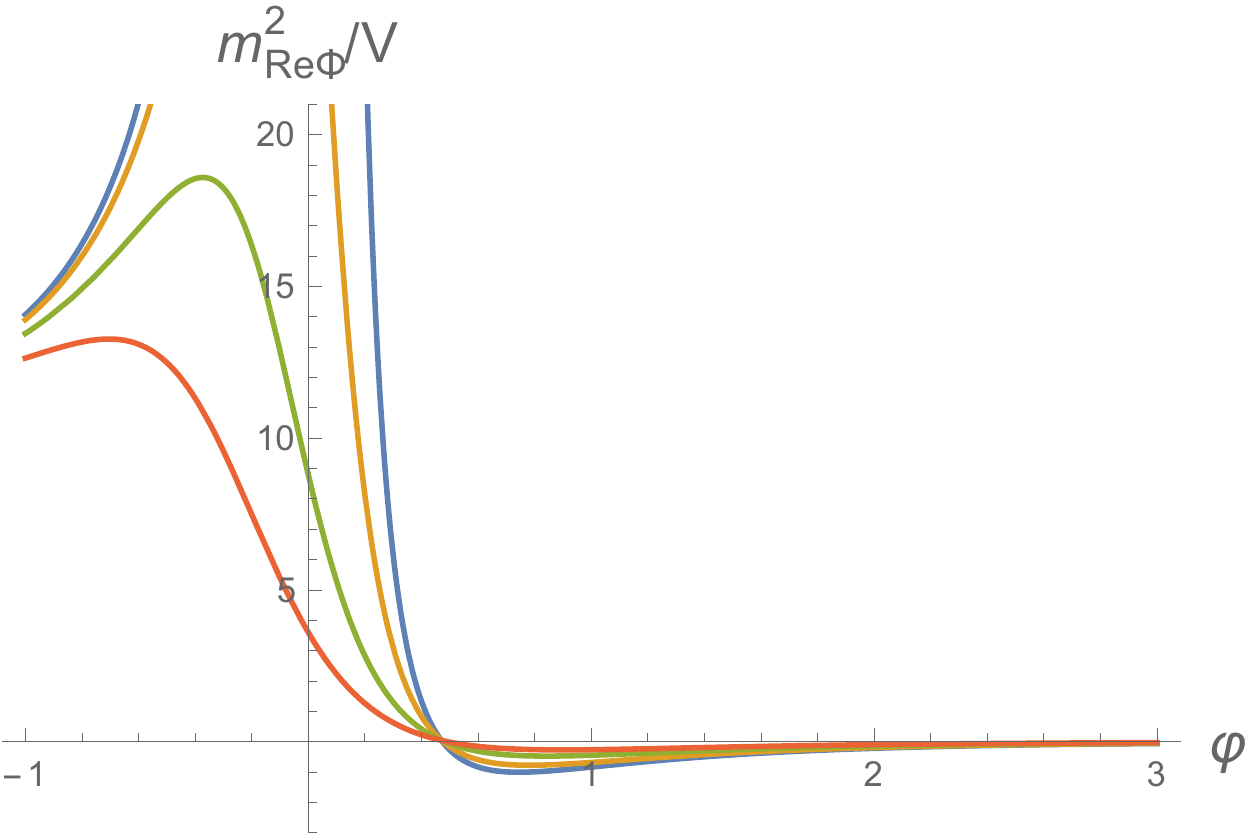}
\includegraphics[width=7.5cm]{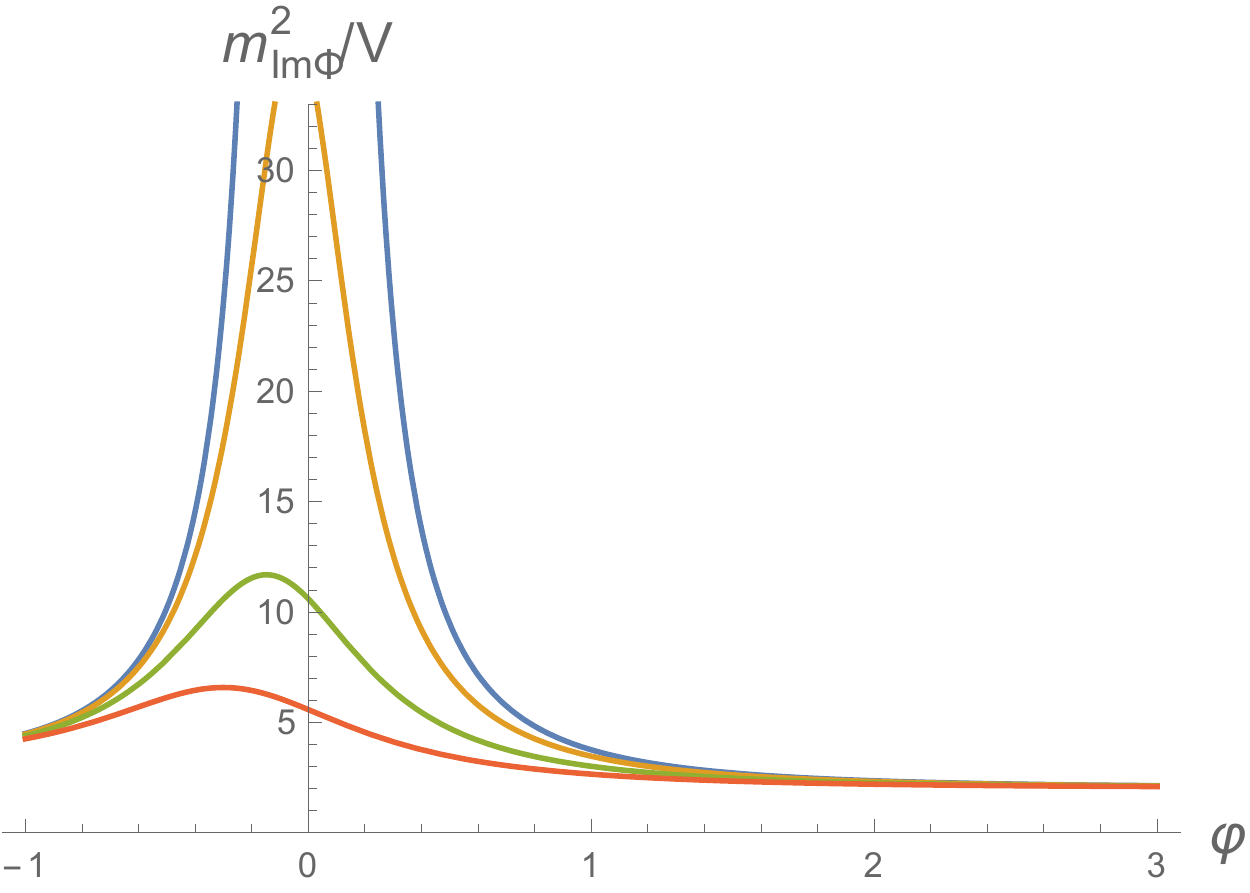}
\caption{Masses of the real and imaginary part of the field $\Phi$ for the model defined by Eq.~\eqref{Example} with $\alpha=1$ and uplifting equal to $\Lambda=\{0, 0.1, 0.3, 0.5\}$. Both scalar parts are massive at the minimum. During inflation, at large values the $\varphi$, the mass of $\Re\Phi$ goes to zero while the mass of $\Im\Phi$ approaches a constant value as defined by Eq.~\eqref{MassImaginary}.}\label{Masses}
\end{center}
\vspace{-0.cm}
\end{figure}

\section{Uplifting geometric  $\alpha$-attractors} \label{GeoUp}

The appealing property of the original formulation of $\alpha$-attractors, as discovered in \cite{Kallosh:2013hoa,Kallosh:2013yoa,Kallosh:2014rga}, is the unique relation between the \Kahler geometry and the observational predictions \eqref{nsandr}. In particular, the logarithmic \Kahler potential fixes the spectral tilt while its constant curvature
\be\label{Kcurv}
R_K=-\frac{2}{3\alpha}\,,
\ee
determines the amount of primordial gravitational waves. However, these original models require always the presence of a second superfield.

Single superfield geometric formulations have been discovered in \cite{Roest:2015qya,Linde:2015uga}. As shown in \cite{Roest:2015qya}, they originates from a natural deformation of the well-known no-scale constructions\footnote{No scale models, as originally proposed in \cite{Cremmer:1983bf,Ellis:1983sf}, represent a good starting point in order to produce consistent inflationary dynamics (see e.g. \cite{Cecotti:1987sa,Ellis:2013xoa,Ellis:2013nxa,Lahanas:2015jwa,Ellis:2015kqa,Ellis:2015pla}). However, the geometric models of this section emerge from a different construction which naturally leads to stable de Sitter solutions and have scale depending on the parameter $\alpha$ (see \cite{Roest:2015qya} for explicit derivation). The no scale symmetry is intimately related to a specific value of the \Kahler curvature \eqref{Kcurv} and it is restored just in the limit $\alpha\rightarrow 1$.}  and they are defined by 
\be\label{alphascaleKW}
K=-3 \alpha \ln\left(\Phi+\bar\Phi\right) \,, \qquad W=\Phi^{n_-} - \Phi^{n_+} F(\Phi)\,,
\ee
with power coefficients equal to
\be\label{npm}
n_\pm = \frac{3}{2} \left(\alpha \pm \sqrt{\alpha}\right)\,,
\ee
and $F$ having general expansion $F(\Phi)=\sum_n c_n \Phi^n$ which encodes the attractor nature of these scenarios.

This class gives rise to the flat $\alpha$-attractors of the previous section in the limit $\alpha\rightarrow\infty$ and, then, when the curvature becomes flat,  as shown in \cite{Roest:2015qya}. The procedure is the following: one performs a field redefinition such as $\Phi \rightarrow \exp(-2 \Phi / \sqrt{3 \alpha})$, an appropriate \Kahler transformation and, in the singular limit, one obtains canonical and shift-symmetric $K$ and $W$ equal to \eqref{fflat}, with F constant. On top of this, one adds exponential corrections which returns the desired inflationary behavior.

In order to uplift the SUSY Minkoswki minimum of these scenarios, one can add a nilpotent field which breaks supersymmetry and yields a non-zero cosmological constant. The geometric analogous of the flat case, discussed in the previous section, is given by

\be\label{GeoUp}
\begin{aligned}
K=-3 \alpha \ln\left(\Phi+\bar\Phi\right) +S\bar{S}\,, \qquad W=\Phi^{\frac{3}{2}\alpha}\left[f(\Phi)+g(\Phi) S \right]\,.
\end{aligned}
\ee
In fact, along the real axis $\Phi=\bar{\Phi}$ and at $S=0$, this supergravity model yields a scalar potential
\be
V=8^{-\alpha}\left[g(\Phi)^2-3f(\Phi)^2+\frac{4 \Phi^2 f'(\Phi)^2}{3\alpha}\right]\,,
\ee
which, when expressed in terms of the canonical field $\varphi=-\sqrt{3\alpha/2}\ln \Phi$, coincides with the one obtained in the flat case Eq.~\eqref{pot}, up to an overall constant factor. Furthermore, Eq.~\eqref{GeoUp} reduces to Eq.~\eqref{K&W} in the flat singular limit. The \Kahler potential \eqref{GeoUp} parametrizes a manifold $SU(2,1)/U(1)\times U(1)$ and  related analysis with similar settings are performed in \cite{Lahanas:2015jwa,CKL}.

The correspondence between the scalar potentials of the flat and the geometric construction (for the single superfield case it was proven in \cite{Roest:2015qya}) is remarkable as it allows to identically assume the whole set of results, from Eq.~\eqref{f&gflat} to Eq.~\eqref{MassImaginary}, found and described in the previous section, provided one identifies
\be
x\equiv\Phi\,.
\ee
The functions $f$ and $g$ can be assumed to have generic expansion \eqref{f&gflat} and the inflationary behavior will be of the form \eqref{falloff}. However, in this case, the fall-off will be governed by the curvature of the \Kahler manifold which depends on the parameter $\alpha$. The minimum, placed at $\Phi=1$, provided
\be
f'(1)=g'(1)=0\,,
\ee
will have uplifting equal to \eqref{CC}, gravitino mass and SUSY breaking scale given by \eqref{M32M} and, again, supersymmetry broken just in the $S$ direction, as given by
\be
D_S W_{min} = g(1) = M\,, \qquad D_{\Phi}W_{min}=0,.
\ee

Remarkably, the condition on the stability of the inflationary trajectory turns out to be the same of the previous section. At large value of the canonical field $\varphi$, the mass of $\Im\Phi$ is positive when Eq.~\eqref{massIm} is satisfied, independently of the value of $\alpha$\footnote{Note added: related results on stability have been derived in the paper \cite{Carrasco:2015rva}, appeared on the same day of submission of the present work.}. This represents a considerable improvement  with respect to the single superfield case defined by \eqref{alphascaleKW} which is stable just for $\alpha>1$ \cite{Roest:2015qya}. Furthermore, the mass of $\Im\Phi$ approaches a constant value during inflation as given by \eqref{MassImaginary}, up to an overall constant.

\section{General inflaton potential from curved \Kahler geometry}

In the previous section, we have developed the general framework in order to obtain inflation together with controllable level of uplifting and SUSY breaking at the minimum when the \Kahler geometry is curved and defined by Eq.~\eqref{GeoUp}. We have proven that generic expansion of $f$ and $g$ gives rise to $\alpha$-attractors with cosmological predictions extremely stable. 

On the other hand, also in this context, it is possible to make the specific choice \eqref{g&f} and consider the geometric analogous of the class of models introduced in \cite{Dall'Agata:2014oka,Kallosh:2014hxa} and reviewed in Sec.~\ref{InflDS}. Then, the \Kahler potential and the superpotential read
\be
\begin{aligned}
K=-3 \alpha \ln\left(\Phi+\bar\Phi\right) +S\bar{S}\,, \qquad W=\Phi^{\frac{3}{2}\alpha} f(\Phi)\left(1+\frac{S}{\beta} \right)\,.
\end{aligned}
\ee
The choice $\beta=1/\sqrt{3}$ gives rise to a scalar potential with a Minkowki minimum. Along $\Phi=\bar{\Phi}$ and $S=0$, one has (up to an overall constant factor)
\be
V=\frac{2 }{3\alpha}\Phi^2 f'(\Phi)^2\,,
\ee
which, in terms of the canonical scalar field $\varphi$ reads
\be
V= f'\left(\e^{-\sqrt{\tfrac{2}{3\alpha}}\varphi}\right)^2\,,
\ee
where primes denote derivatives with respect to the variables the function depends on. Then, one can implement an arbitrary inflaton potential, independently of the value of the \Kahler curvature which is parametrised by $\alpha$. Related results for the case $\alpha=1$ were obtained in \cite{Lahanas:2015jwa}. In the case of a flat \Kahler geometry the works \cite{Kallosh:2010xz,Dall'Agata:2014oka,Kallosh:2014hxa} developed analogous constructions.

Within this setup, one can implement even a quadratic potential $V=\tfrac{1}{2}m^2\varphi^2$ by choosing
\be
f(\Phi)=\frac{3\alpha\ m}{4\sqrt{2}}\ln^2(\Phi)\,.
\ee

The properties at the minimum remain the same as in the flat case of Sec.~\ref{InflDS}. Then, a small deviation of $\beta$ from the value $1/\sqrt{3}$ yields the desirable tiny uplifting which reproduces the current acceleration of the Universe.

\section{Discussion}\label{disc}

In this paper, we have provided evidences for the special role that  $\alpha$-attractors would play in the  cosmological evolution of the Universe. In the simple supergravity framework consisting of two sectors (one containing the inflaton and the other controlling the landscape of possible vacua), any arbitrary expansion of the superpotential would yield automatically such inflationary scenarios. We have obtained these results both in the case of a flat \Kahler geometry, as given by Eq.~\eqref{K&W}, and in the case of the logarithmic \Kahler as defined by Eq.~\eqref{GeoUp} where the geometric properties of the \Kahler manifold determines the observational predictions. In this latter case, the overall factor $\Phi^{\frac{3}{2}\alpha}$ in $W$ can be removed by means of an appropriate \Kahler transformation (this choice makes the shift symmetry of the canonical inflaton $\varphi$ manifest even in the case of a logarithmic \Kahler potential, as pointed out in \cite{Carrasco:2015uma}). However, one would lose immediate contact with string theory scenarios as the form of $K$ would change consequently. In this respect, polynomial contributions to the superpotential, typically arising from flux compactification, would be possible if 
\be
\alpha=\frac{2}{3}n
\ee
with $n$ integer. In particular, the simple choice $n=1$ would give
\be
\begin{aligned}
K&=-2\ln\left(\Phi+\bar\Phi\right) +S\bar{S}\,,\\
 W&=\left(a_0\Phi + a_1\Phi^2 +...\right)+ \left(b_0\Phi + b_1\Phi^2 +...\right) S \,,
\end{aligned}
\ee
where dots stand for higher order terms in $\Phi$ (see \cite{Dudas:2015lga} for a recent analysis of this class of models in the context of supplementary moduli breaking supersymmetry). Then, the minimal addition of a nilpotent sector with canonical $K$ to the class proposed in \cite{Roest:2015qya} leads to a simplification of the original superpotential \eqref{alphascaleKW} and enhancement of stability of the inflationary trajectory, which now occurs for any value of $\alpha$ (see \cite{Carrasco:2015uma} for a discussion on the connection between curvature and stabilization).

We have shown that  cosmological $\alpha$-attractors are absolutely insensitive with respect to any value of the cosmological constant and to the coupling between $\Phi$ and $S$. The plateau and the fall-off turn out to be extremely stable with respect to generic deformations of the superpotential (similar stability can be observed in some examples of \cite{Kallosh:2015lwa}). These scenarios would arise naturally in any possible Universe, independently of the amount of dark energy. In this regard, cosmological attractors seem to be fundamentally compatible with the idea of Multiverse and landscape of vacua.

Quantum corrections or interactions with other particles may lead to some additional contributions. However, this should not affect the existence of a landscape of dS vacua and any possible correction to the cosmological constant would be easily faced, within such a scenario with controllable level of dark energy. Furthermore, while higher-order corrections may affect the inflationary regime (one generically gains a better control in the context of $F(R)$ supergravity \cite{Ketov:2012yz}), we have discussed the general implications of combining the inflaton with a nilpotent field focusing just on the lower-order effects. Proper string theory realizations may shed additional light on this phenomenon (see \cite{Kallosh:2015nia} for a recent study on the topic of the nilpotent field in string theory).

The cosmological predictions \eqref{nsandr} of this class of models, being in excellent agreement with the latest Planck data \cite{Planck:2015xua,Ade:2015lrj}, belong to a region in the $(n_s,r)$ plane which seems to be very appealing from the theoretical viewpoint. Higher order terms of the Ricci scalar in a pure gravitational Lagrangian, such as the Starobinsky model \cite{Starobinsky:1980te} and its supergravity realizations \cite{Cecotti:1987sa, Ellis:2013xoa,Kallosh:2013lkr,Buchmuller:2013zfa,Farakos:2013cqa,Ellis:2013nxa}, lead to \eqref{nsandr} with $\alpha=1$ (by including an auxiliary vector field one can vary the value of $\alpha$ \cite{Ozkan:2015iva}). Further, models with non-minimal couplings, such as Higgs inflation \cite{Bezrukov:2007ep} and the universal attractor model \cite{Kallosh:2013tua}, yield identical observational predictions. Interestingly, the peculiarity of such a region translates into a common denominator being a pole of order two in the kinetic term of the inflaton \cite{Galante:2014ifa}. Finally, recent studies on the excursion of the inflaton field have revealed a change of its behavior just around the region defined by \eqref{nsandr} \cite{Garcia-Bellido:2014eva,Garcia-Bellido:2014wfa}. It would be of great interest to go further with the search and try to understand whether something more fundamental still needs to be unveiled.



\section*{Acknowledgments}
MS would like to thank Renata Kallosh, Andrei Linde and Diederik Roest for enlightening discussions and precious comments on a preliminary version of the paper. Further, MS acknowledges stimulating discussions with Michele Cicoli, Pablo Ortiz, Mehmet Ozkan, Fernando Quevedo, Pelle Werkman and Clemens Wieck.

\bibliography{refsUS}
\bibliographystyle{JHEP}

\end{document}